\documentclass{article}
\usepackage{
amsmath,
amssymb,
rotating,
times,
alltt,
}

\DeclareGraphicsExtensions{.eps} 

\begin{document}

\title{Motif Discovery through Predictive Modeling of Gene Regulation}
\newcommand{\URL}{medusa}

\author{
Manuel Middendorf$^{1}$,
Anshul Kundaje$^{2}$ ,
Mihir Shah$^{2}$,
Yoav Freund$^{2,4,5}$,\\
Chris H. Wiggins$^{3,4}$,
and
Christina Leslie$^{2,4,5}$
\\\smallskip\\
$^1$ Department of Physics,\\
$^2$ Department of Computer Science,\\
$^3$ Department of Applied Mathematics,\\
$^4$ Center for Computational Biology and Bioinformatics,\\
$^5$ Center for Computational Learning Systems\\
Columbia University, New York, NY 10027,\\
{cleslie@cs.columbia.edu}\\
{http://www.cs.columbia.edu/compbio/\URL}
}
%\authorrunning{M. Middendorf {\em et al.}}

\maketitle

\begin{abstract}
We present MEDUSA, an integrative method for learning motif models of transcription factor binding sites by incorporating promoter sequence and gene expression data.  We use a modern large-margin machine learning approach, based on boosting, to enable feature selection from the high-dimensional search space of candidate binding sequences while avoiding overfitting.  At each iteration of the algorithm, MEDUSA builds a motif model whose presence in the promoter region of a gene, coupled with activity of a regulator in an experiment, is predictive of differential expression.  In this way, we learn motifs that are functional and predictive of regulatory response rather than motifs that are simply overrepresented in promoter sequences.  Moreover, MEDUSA produces a model of the transcriptional control logic that can predict the expression of any gene in the organism, given the sequence of the promoter region of the target gene and the expression state of a set of known or putative transcription factors and signaling molecules.  Each motif model is either a $k$-length sequence, a dimer, or a PSSM that is built by agglomerative probabilistic clustering of sequences with similar boosting loss.  By applying MEDUSA to a set of environmental stress response expression data in yeast, we learn motifs whose ability to predict differential expression of target genes outperforms motifs from the TRANSFAC dataset and from a previously published candidate set of PSSMs.  We also show that MEDUSA retrieves many experimentally confirmed binding sites associated with environmental stress response from the literature.
\end{abstract}

\section{Introduction}
One of the central challenges in computational biology is
the elucidation of mechanisms for gene transcriptional regulation
using functional genomic data.
The problem of identifying
binding sites for transcription factors
in the regulatory sequences of genes 
is a key component in these computational efforts.
While there is a vast literature on this subject, 
only a few different conceptual approaches have been tried, and 
each of these standard approaches has its limitations.

The most widely-used methodology
for computational discovery of putative binding sites is based on
clustering genes---usually by similarity of
gene expression profiles, sometimes combined with 
annotation data---and
 searching for motif patterns that are 
overrepresented in the promoter sequences of these genes in the belief
that they may be coregulated.  Popular
motif discovery programs 
in
this paradigm include MEME \cite{bailey:fitting},
Consensus \cite{hertz:identifying}, Gibbs Sampler \cite{lawrence:detecting}, AlignACE \cite{hughes:alignace} and many others.  
The cluster-first methodology has several drawbacks.  First,
it is not always true that genes with correlated gene 
expression profiles are in fact coregulated genes
whose regulatory regions contain common binding sites.  
Moreover, by focusing on 
coregulated genes, one
fails to consider more complicated combinatorial regulatory programs
and the overlapping regulatory pathways that can affect different sets
of genes under different conditions.  Recently, more sophisticated 
graphical models for gene expression data 
have been introduced to try to partition genes into
``transcriptional modules'' \cite{segal:module}---clusters of genes that obey a common 
transcriptional program depending on a small number of 
regulators---or to learn overlapping clusters of this kind \cite{battle:overlapping}.  These graphical model approaches use the abstraction of modules to give an interpretable representation of putative relationships between genes 
and to suggest biological hypotheses.  One expects that
using these more complex clustering algorithms
as a preprocessing step for motif discovery would lead to improved
identification of true binding sites; however, it is difficult 
to assess how much of an advantage one might obtain.

Another well-established motif discovery approach is the 
innovative
REDUCE 
method \cite{bussemaker:reduce} and related algorithms \cite{conlon:regression,zilberstein:regression}.  REDUCE avoids the cluster-first methodology by
considering the genome-wide expression levels given by a single
microarray experiment, and it discovers sequences whose presence
in promoter sequences correlates with differential expression.  
Since REDUCE uses linear regression to iteratively
identify putative binding sites, it must enforce strict
tests of statistical significance to avoid overfitting in a large
parameter space corresponding to the set of all possible sequence
candidates.  Therefore, REDUCE can find the strongest signals in
a dataset but will not attempt to find more subtle sites that affect
fewer genes.  
Since the algorithm fits parameters independently 
for each microarray experiment, the issue of
condition-specific regulation enters the analysis
only as post-processing step rather than through
simultaneous training from multiple conditions.

In this paper, we introduce a new motif discovery algorithm called
MEDUSA (Motif Element Discrimination Using Sequence Agglomeration) that learns putative binding sites associated with condition-specific
regulation in a large gene expression dataset.  
MEDUSA works by extracting binding site
motifs that contribute to a 
\emph{predictive model} of gene regulation.  
More specifically, MEDUSA builds  
motif models whose presence in the promoter
region of a gene, together with the
activity of regulators in an experiment,
is predictive of differential expression.
Like REDUCE,
MEDUSA avoids the cluster-first methodology
and builds a single regulatory model to explain the response
of all target genes.  However, unlike REDUCE, MEDUSA learns from
multiple and diverse gene expression experiments, using the
expression states of a set of known regulatory to represent
condition-specific regulatory conditions.
Moreover, MEDUSA 
is based on a classification approach (using
large-margin machine learning)
rather than linear regression, 
to avoid overfitting in the high-dimensional search
space of candidate binding sequences.
In addition to discovering binding site motifs, MEDUSA produces a model of
the condition-specific transcriptional control logic 
that can predict the expression of
any gene, given 
the gene's promoter sequence 
and 
the expression state of a set of known transcription factors and signaling molecules.

The core of 
MEDUSA is a boosting algorithm that
adds a binding site motif 
(coupled with a regulator whose
activity helps predict up/down regulation of genes whose promoters contain
the motif) 
to an overall gene regulation model at each boosting iteration.  
Each motif model is 
either a $k$-length sequence (or ``$k$-mer''),
a dimer, or a PSSM.  The PSSMs are generated by considering
the 
most predictive $k$-mer features (Fig.~\ref{fig:code})
selected at a given round of boosting
that are associated with a common regulator; we 
then
perform
agglomerative
probabilistic clustering of these $k$-mers into PSSMs, and we select
from all the candidate PSSMs seen during clustering the one that 
minimizes boosting loss (Fig.~\ref{fig:code}).  In experiments on 
a set of environmental stress response
expression data in yeast, we learn motifs together with  regulation
models that achieve accurate prediction of up/down regulation of target
genes in held-out experiments.  In fact, we show that
the 
performance 
 of the learned motifs for prediction of 
differential expression 
in test data
is stronger 
than the performance of  motifs 
from the TRANSFAC dataset or from a previously published
candidate set of PSSMs.  For these environmental stress response 
experiments, we also show that 
MEDUSA retrieves many experimentally confirmed binding sites from the 
literature.

We first introduced the idea of \emph{predictive modeling} of gene
regulation with the GeneClass algorithm \cite{middendorf:geneclass}.  However, GeneClass
uses a fixed set of candidate motifs as an input to the algorithm and
cannot perform motif discovery.  We note also that there have been previous
efforts to incorporate motif discovery in an integrative model for
sequence and expression data using the probabilistic graphical model framework \cite{segal:learningmod}.  This graphical model approach again uses the abstraction of ``modules'' to learn sets of motifs associated with clusters of genes, giving a high-level modular representation of gene regulation.  As explained above, MEDUSA does not produce an abstract module representation.  However,
it has two advantages over graphical model methods.  First, MEDUSA uses
a large-margin learning approach that helps to improve the 
\emph{generalization}
of the learned motifs and regulation model, and we can evaluate
prediction accuracy on held-out 
experiments to assess our confidence in the model.  Second, training
graphical models requires special expertise to avoid 
poor
local minima
in a complex optimization problem, while MEDUSA can be run ``out-of-the-box''.
Code for MEDUSA is publicly available and can be downloaded
from the supplementary website for the paper,
http://www.cs.columbia.edu/compbio/medusa.

\section{Methods}

\subsection{Learning Algorithm}
MEDUSA learns binding site motifs together with 
a predictive gene regulation model
using a specific implementation of
Adaboost, a general
discriminative learning algorithm proposed by Freund 
and Schapire~\cite{Schapire02}. Adaboost's
basic idea is to 
iteratively apply variants of a simple, weakly discriminative
learning algorithm,
called
the {\em weak} learner, to different weightings of the same
training set. 
The only requirement of the weak learner is that it predicts
the class label of interest
with greater than 50\% accuracy.
At each iteration, weights are recalculated so that examples
which were misclassified at the previous iteration are more
highly weighted.
Finally, all of the weak prediction rules are combined into 
a single {\em strong} rule using a weighted majority vote.
As discussed in~\cite{SchapireFrBaLe98},
boosting is a large-margin classification algorithm,
able to learn from a potentially large number of candidate features
while maintaining good generalization error (that is,
without over-fitting the training data).

The discretization of expression data (see Sect.~\ref{sec:disc}) into
up- and down-regulated expression levels allows us
to formulate the problem of predicting regulatory response
of target genes as the \emph{binary classification} task of learning
to predict up and down examples.
Rather than 
viewing each microarray experiment as a training example,
MEDUSA considers all genes and experiments simultaneously and
treats every gene-experiment pair as a separate instance, 
dramatically increasing the 
number of training examples
available.
For every gene-experiment example, the gene's expression state in the
experiment
(up- or down-regulation) gives the output label $y_{ge}=\pm$.
As we explain below (see Sect.~\ref{sec:disc}), positive and negative examples correspond to statistically significant up- and down-regulated expression levels; examples with baseline expression levels are omitted from training.

The inputs to the learner are (i) the promoter sequences 
of the target genes and (ii) the discretized expression levels of a set of 
putative regulator genes.
The sequence data is represented only via occurrence or non-occurrence
of a sequence element or motif.  A full discussion of how MEDUSA determines
a set of sequence and motif candidates to be considered at each round
of boosting is given in Sect.~\ref{sec:HC}.
Let the binary matrix $M_{\mu g}$ indicate the presence ($M_{\mu g}=1$) or
absence ($M_{\mu g}=0$) of a motif $\mu$ in the 
promoter sequence of gene $g$,
and let the binary matrices 
$P_{\pi e}^{\sigma}$ indicate the up-regulation ($\sigma=+$)
or down-regulation ($\sigma=-$) of a regulator $\pi$ in
experiment $e$ ($P_{\pi e}^{\sigma}=1$, if regulator $\pi$
is in state $\sigma$ in experiment $e$, and $P_{\pi e}^{\sigma}=0$,
otherwise).
Our weak rules split the gene-experiment examples in the training data
by asking questions of the form \mbox{`$M_{\mu g}P_{\pi e}^{\sigma}=1$ ?';}
i.e., `Is motif $\mu$ present, and is regulator $\pi$ in state $\sigma$?'.
In this way, each rule introduced corresponds to a putative 
interaction between a regulator and some sequence element in the promoter 
of the target gene that it regulates.

\begin{figure}[htb]
\begin{center}
\includegraphics[width=2.5in]{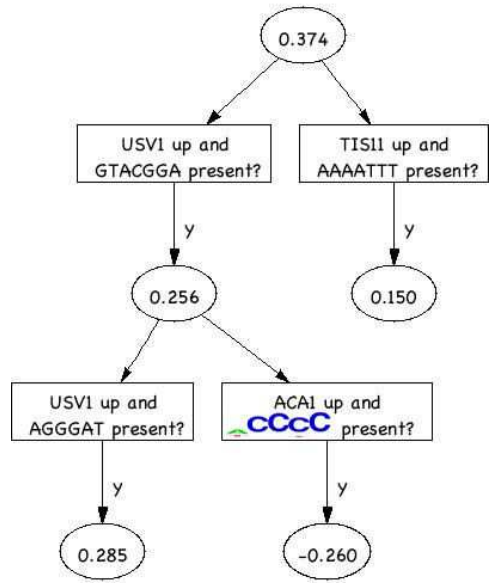}
\end{center}
\caption{\footnotesize {\bf Example of an alternating decision tree:} The rectangles
represent 
weak rules, learned by MEDUSA,
that split gene-experiment examples in the training data.
Examples for which the condition holds follow the path further down the tree (`y') and
have their scores incremented by the prediction score given in the ovals. The final prediction is the sum of
all scores that the example reaches.
\label{fig:adt}}
\end{figure}

The weak rules are combined by weighted majority vote using 
the structure of an alternating decision tree~\cite{freund:alt,middendorf:geneclass}.
An example is given in Fig.~\ref{fig:adt}. 
The weak rules are shown in rectangles. Their associated
weights, indicating the strength of their contribution to the
majority vote, 
are shown in ovals. 
If the \{motif presence, regulator state\} condition for a particular rule
holds 
in the example considered,
the weight
of the rule is added to the final prediction score. The weight
can be either positive or negative, contributing to up- or
down-regulation respectively. Rules that appear lower in the
tree are conditionally dependent on the rules in ancestor nodes. For
example, in Fig.~\ref{fig:adt},
only if USV1 is up-regulated and both motifs GTACGGA
and AGGGAT are present is the score 0.285 added to the prediction
score. The tree structure is thus able to reveal combinatorial
interactions between regulators and/or motifs. 
The sign of the final prediction score gives the prediction, and
the absolute value of the score indicates the level of confidence. 
In this work, we consider both sequences and 
position-specific scoring matrices (PSSM) (an example is shown in the lower right
node of Fig.~\ref{fig:adt}) as putative motifs (see Sect.~\ref{sec:HC}). 

Each iteration of the boosting algorithm results in the addition of a new
node (corresponding to a new weak rule) to the tree.
The weak rule and its position in the tree at which it is added are chosen by minimizing the boosting
loss over all possible combinations of motifs, regulators, 
and regulator-states, and over all possible positions (``preconditions'')
 in the current tree.
A pseudo-code description  
is given in Fig.~\ref{fig:code}.

The implementation uses efficient sparse matrix multiplication in MATLAB, 
exploiting the fact that our motif-regulator features are outer products
of motif occurrence vectors and regulator expression vectors, and
allows us to scale up to significantly larger datasets than in~\cite{middendorf:geneclass}.

\begin{figure}[!h]
\begin{center}
\begin{tabular}{|l|}
\hline
{\em Definitions:}\\
\begin{tabular}{lcp{1.7in}}
$\hat{c}$ &=&precondition associated with a specific 
position 
in the tree\\
$c_{\mu\pi\sigma}$ &=&weak rule associated with motif $\mu$
and regulator $\pi$ in state $\sigma$\\
$w_{ge}$ &=&weight of example ($g,e$)\\
$W[c(g,e)]$&=&$\sum_{c(g,e)=1}w_{ge}$, for a given condition $c$\\
$\lnot{c}$&=&not $c$\\
$Z(\hat{c},\mu,\pi,\sigma)$&=&boosting loss\\
\multicolumn{3}{r}{= $W[\lnot \hat{c}]
+2\sqrt{W[\hat{c}\land c_{\mu\pi\sigma}]W[\hat{c}\land\lnot c_{\mu\pi\sigma}]}$}\\
$y_{ge}$&=&label of example ($g,e$)\\
$T$&=&total number of boosting iterations\\
$F_t(g,e)$&=&prediction function at iteration $t$\\
$\alpha_t$&=&weight of weak rule $t$ contributing to 
the final prediction score\\
\end{tabular}\\
{\em Initialization:}\\
$F_0(g,e) = 0$, for all ($g,e$)\\
{\em Main loop:}\\
for $t=1\ldots T$\\
\hspace*{.4cm}$w_{ge}=e^{-y_{ge}F_{t-1}(g,e)}$\\
\hspace*{.4cm}call Hierarchical Motif Clustering (Sec.~\ref{sec:HC}).\\
\hspace*{.4cm}get a set of proposed PSSMs.\\
\hspace*{.4cm}minimize boosting loss:\\
\hspace*{.4cm}$\mathbf{c}^*=\mbox{argmin}_{\hat{c},\mu,\pi,\sigma}Z(\hat{c},\mu,\pi,\sigma)$\\
\hspace*{.4cm}calculate weight of the new weak rule $\mathbf{c}^*$:\\
\hspace*{.4cm}$\alpha_t = \frac{1}{2}\ln{\frac{W[\mathbf{c}^*\land (y_{ge}=+)]}{W[\mathbf{c}^*\land (y_{ge}=-)]}}$\\
\hspace*{.4cm}add new node $\mathbf{c}^*$ with weight $\alpha_t$ to the tree\\
\hspace*{.4cm}$F_t(g,e) = F_{t-1}(g,e) + \alpha_t\mathbf{c}^*(g,e)$\\
end for\\
$\mbox{sign}(F_T(g,e)) =$ prediction for example $(g,e)$\\
$|F_T(g,e)| = $ prediction confidence for $(g,e)$\\
\hline
\end{tabular}
\end{center}
\caption{{\footnotesize {\bf Pseudo-code description of the learning algorithm}}}
\label{fig:code}
\end{figure}

\subsection{Hierarchical Motif Clustering}
\label{sec:HC}
At each boosting iteration, MEDUSA considers all occurrences of $k$-mers ($k=2,3,\ldots 7$) and dimers
with a gap of up to 15 bp (see Sect.~\ref{sec:motifset}) in the promoter
sequence of each gene as candidate motifs.  Since slightly different
sequences might in fact be instances of binding sites for the same regulator, 
MEDUSA performs a hierarchical motif clustering algorithm to generate
more general candidate PSSMs as binding site models.
The motif clustering
uses \mbox{$k$-mers} and dimers associated with low boosting loss as a starting point
to build PSSMs: these sequences are viewed seed PSSMs, and then
the algorithm proceeds by iteratively merging similar PSSMs, as described
below. 
The generated
PSSMs are then considered as additional putative motifs for the learning
algorithm.

A position-specific scoring matrix (PSSM)
 is represented by 
a probability distribution $p(x_1,x_2,\ldots,x_n)$ over sequences
$x_1x_2\ldots x_n$, where $x_i\in\{A,C,G,T\}$. The emission probabilities
are assumed to be independent at every position such that 
$p(x_1,\ldots,x_n)=\prod_{i=1}^n{p_i(x_i)}$. For a given input sequence
the PSSM returns a log-odds score 
$S = \sum_{i=1}^n\ln{({p_i(x_i)}/{p^{bg}(x_i)})}$
with respect to background 
probabilities $p^{bg}$. A score threshold can then be chosen to define whether
the input sequence is a hit or not.

When comparing two PSSMs, we allow possible offsets between the two starting positions.
In order to give them the same lengths, we pad either
the left or right ends with the background distribution.
We then define
a distance measure $d(p,q)$ as the minimum over all possible position offsets 
of the JS entropy~\cite{cover:infobook} between two PSSMs $p$ and $q$. 
\begin{equation*}
d(p,q)\equiv \min_{\rm offsets}\big[w_1D_{KL}(p||w_1p+w_2q) 
+  w_2 D_{KL}(q||w_1p+w_2q)\big],\\
\end{equation*}
where $D_{KL}$
is the Kullback-Leibler divergence~\cite{cover:infobook}. 
By using
$p(x_1\ldots x_n)=\prod_{i=1}^n{p_i(x_i)}$ and
$\sum_{x_i}p_i(x_i)=1$ (and the analogous equations for $q$) one
can easily show that $D_{KL}(p||q)=\sum_{i=1}^nD_{KL}(p_i||q_i)$.
The relative weights of the two PSSMs,
$w_1$ and $w_2$, 
are here defined as $w_{1,2}={N_{1,2}}/{(N_1+N_2)}$, where
$N_1$, $N_2$ are the numbers of target genes for the given PSSM.
Note that this distortion measure is not affected by
adding more ``padded" background elements either before or after the PSSM.
Our merge criterion is similar to the one used in the
agglomerative information bottleneck algorithm \cite{slonim:aib}, 
though we also consider offsets in our merges.

At every boosting iteration, we first find the weak rule $c_{tmp}$
among all possible combinations of regulators, regulator-states and
sequence motifs ($k$-mers and dimers), that minimizes
boosting loss. The
100 motifs with lowest loss appearing with the same regulator, regulator-state,
and precondition as in $c_{tmp}$ are then input to the hierarchical
clustering algorithm. Sequence motifs can be regarded as PSSMs with 0/1 emission
probabilities, smoothed by background probabilities. 
By iteratively joining the PSSMs with smallest $d(p,q)$,
the clustering proposes a set of 99 PSSMs from various stages
of the hierarchy.  
At every merge of two PSSMs, the score threshold associated with the new PSSM
is found by optimizing the boosting loss. 
Note also that the new PSSM can be longer than either of the two PSSMs used
in the merge, due to the procedure of merging with offsets; in this way, we
can obtain candidate PSSMs longer the maximum seed $k$-mer length of $7$.
The number of target genes,
which determines the weight of the PSSM for further clustering, is 
calculated by counting the number of promoter sequences 
which score above the threshold.
The new node that
is 
then added to the alternating decision tree 
is the weak rule that minimizes boosting
loss considering all sequence motifs and PSSMs.

\section{Statistical Validation}
\subsection{Dataset}
We use the environmental
stress response (ESR) dataset of 
Gasch {\em et al.}~\cite{gasch:yeast},
which consists of 173
cDNA microarray experiments measuring the expression of 6152 
{\em S. cerevisiae} genes in response to diverse environmental
perturbations.
All measurements are given as 
$\log_2$ expression values (fold-change with
respect to
an unstimulated reference 
condition).
Note that our analysis does not
require a normalization to a zero-mean,
unit-variance distribution, as is often
employed; instead we wish to retain
the meaning of the true zero (that is, the reference state).

\subsection{Discretization}
\label{sec:disc}
\begin{figure}[htb]
\begin{center}
\includegraphics[scale=0.2]{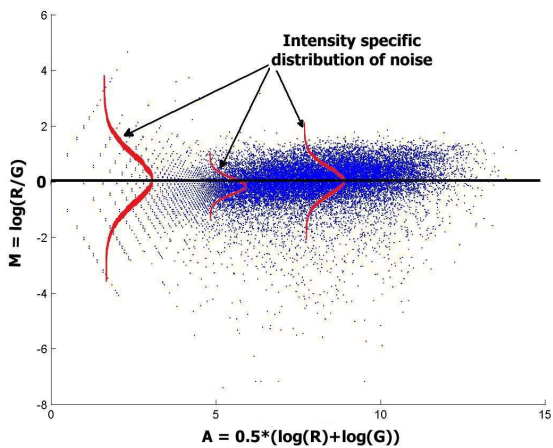}
\caption{\footnotesize {\bf Expression discretization.}  A noise
distribution is empirically estimated using data from three 
unstimulated reference experiments. The noise model takes into
account intensity-specific effects. By choosing a $p$-value cutoff
of 0.05 we discretize differential expression 
into up-regulated, down-regulated, and baseline levels.
}\label{fig:noise}
\end{center}
\end{figure}
We discretize expression data by using a
 noise model that accounts for intensity specific
effects in the raw data from both the Cy3 (R) and Cy5 (G)
channels. In order to estimate the null model, we use 
the three replicate unstimulated experiments published
with the same dataset~\cite{gasch:yeast}.
Plots of $M = \log_{2}({R}/{G})$ versus $A = \log_{2}(\sqrt{RG})$ 
(Fig.~\ref{fig:noise}) show the
intensity specific distribution of the noise in the expression values.
We compute the cumulative
empirical null distribution of $M$ conditioned on $A$ by binning the $A$
variable into small bin sizes, 
maintaining a good resolution
while having sufficient data points per bin. 
For any
expression value $(M,A)$ of a gene in an experiment, we estimate
a $p$-value based on the null distribution conditioned on $A$,
and we
use a $p$-value cutoff of 0.05 to discretize the expression values
into  +1, -1 or 0 (up-regulation, down-regulation, or baseline).
The discretization allows us to formulate the prediction problem
as a classification task.

\subsection{Candidate Regulators}
The regulator set consists of 475 genes
(transcription factors, signaling
molecules, kinases and phosphatases), including
466 which are used in Segal
{\em et al.}
\cite{segal:module} and 9 generic (global) regulators obtained
from Lee {\em et al.}~\cite{lee:binding}.

\subsection{Motif Set}
\label{sec:motifset}
We scan the 500 bp 5'-UTR promoter
sequences of all \emph{S. cerevisiae} genes from the Saccharomyces genome
Database (SGD) for all occurring $k$-mer motifs ($k=2,3,\ldots, 7$).
We also include 3-3 and 4-4 dimer motifs allowing a middle gap
of up to 15 bp.  We restrict the
set of all dimers to 
those whose two components have 
specific relationships, consistent with most known dimer motifs:
equal, reversed, complements, or reverse-complements.
As described in Sect.~\ref{sec:HC}, we use an information-theoretic,
hierarchical clustering scheme 
to 
infer
a set of PSSMs at each boosting iteration. 
The complete candidate
motif set is 
then 
the union of all \mbox{$k$-mers}, dimers, and PSSMs, 
with a cardinality of \mbox{$10962+1184+99=12245$}.

\subsection{Cross-validation}

We divide the 173 microarray experiments into five folds,
keeping replicate experiments in the same fold. 
We then perform five-fold cross-validation, training the classifier
on four folds and testing it on the held-out fold. The learning
algorithm is run for 700 boosting iterations.
The average test-loss for prediction
on all genes in held-out experiments is $13.4\pm 3.9\%$.

For comparison, we run the same learning algorithm with
experimentally-confirmed or computationally-predicted motifs
in the literature. In these runs, the hierarchical motif clustering
is left out, and the set of putative motifs contains
only those that were proposed in the literature. 

The TRANSFAC database~\cite{wigender:transfac} contains a library
of known and putative binding sites which can be used to scan
the promoter sequence of every gene. After removing redundant
sites, we compile a list of 354 motifs. The boosting algorithm
with the same number of iterations and the same folds for 
cross-validation gives a higher test-loss of $20.8\pm 2.8\%$
The
compiled TRANSFAC motifs thus have a much weaker
strength in predicting gene expression than the motifs
found by MEDUSA.

The same comparison was performed with a list of 356 motifs
found in~\cite{pilpel:motifsyn} by using a state-of-the-art Gibbs
sampling algorithm on groups of genes clustered by expression data
and annotation information.
These motifs also gave weaker predictive strength than 
those discovered by MEDUSA with an average test-loss of
$16.1\pm 3.5\%$.

We are thus able to identify motifs which have a significantly
stronger prediction accuracy (on independent held-out experiments)
than motifs previously identified in the literature.

\section{Biological Validation}

To confirm that MEDUSA can retrieve biologically meaningful motifs, we
run additional experiments, randomly holding out $10\%$ of the 
(gene,experiment) examples and training MEDUSA on the
remaining examples. We learn ungapped $k$-mers and dimers
simultaneously. After 1000 iterations, we obtain a test loss of
$11\%$ and a set of 1000 PSSMs. We then compare to several known
and putative binding sites, consensus sequences and PSSMs from five
databases: TRANSFAC~\cite{wigender:transfac}, TFD, SCPD, YPD
and a set of PSSMs found by AlignACE~\cite{pilpel:motifsyn}. After
converting the sequences and consensus patterns to PSSMs, smoothed
by background probabilities, we compare all PSSMs with the ones
found by MEDUSA using $d(p,q)$ (see Sect.~\ref{sec:HC}) as a
distance measure. We define the best match for each of MEDUSA's
PSSMs as the PSSM that is closest to it in terms of $d(p,q)$.

\begin{figure*}[htb]
\begin{center}
\includegraphics[height=6.5in]{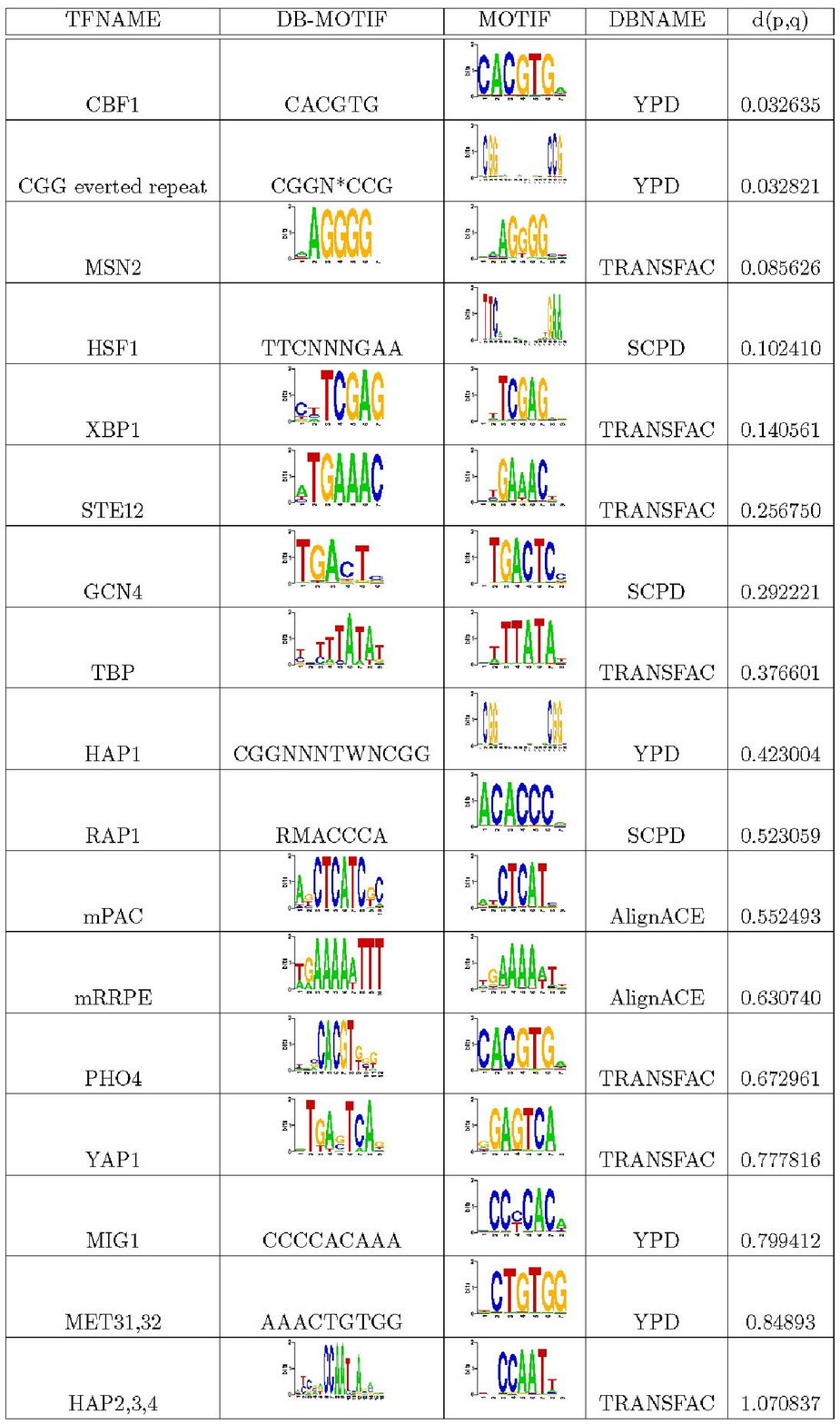}
\end{center}
\caption{\footnotesize {\bf Matching MEDUSA's PSSMs to motifs known
in the literature:} By using $d(p,q)$ (see Sect.~\ref{sec:HC})
as a distance measure, we match PSSMs identified by MEDUSA's to
motifs known in the literature. The table shows the logos of MEDUSA's
PSSMs (column 3), the matching motif of the database (column 2),
the corresponding transcription factor (column 1), the name of the
database (column 4) and the distance $d(p,q)$ (column 5).
\label{fig:pssm}}
\end{figure*}

Each node in the alternating decision tree defines a particular
subset of genes, namely those having at least one example that passes
through the particular node. In this way, we can associate motifs
with Gene Ontology (GO) annotations by looking for enriched GO
annotations in the gene subsets, and we can estimate the
putative functions of the targets of a transcription factor that
might bind to the PSSM in each node. We see matches to variants of
the STRE element, the binding site for the MSN2 and MSN4 general
stress response transcription factors. The genes passing through
nodes containing these PSSMs are significantly enriched for the GO
terms carbohydrate metabolism, response to stress and energy
pathways, consistent with the known functions of MSN2/4. GCR1 and
RAP1 are known to transcriptionally regulate ribosomal genes,
consistent with enriched GO annotations associated with the nodes
of the specific PSSMs. The heat shock factor HSF1---which binds to
the heat shock element (HSE)---plays a primary role in stress
response to heat as well as several other stresses. The heat shock
element exists as a palindromic sequence of the form
\emph{NGAANNTTCN}. We find almost an exact HSE in the tree. In {\em S.
cerevisiae}, several important responses to oxidative and redox
stresses are regulated by Yap1p, which binds to the YRE element.
We find several strongly matching variants of the YRE. It is
interesting to note that comparison of PSSMs from AlignACE with
our PSSMs revealed the PAC and RRPE motifs to be among the top three
matches. These PSSMs also appear in the top 10 iterations in the
tree, indicating they are also strongly predictive of the target
gene expression. Both these putative regulatory motifs have been
studied in great depth with respect to their roles in rRNA
processing and transcription as well their combinatorial
interactions. The enriched GO annotations of these nodes are the
same as their putative functions. The tree contains 122 dimer
motifs with variable gaps. These include the HSE motif
(\emph{GAANNNTTC}), HAP1 motif (\emph{CCGN*CCG}), GIS1 motif
(\emph{AGGGGCCCCT}) as well as variants of the \emph{CCG} everted
repeat. Several important biologically verified PSSMs learned by
MEDUSA are given in Fig.~\ref{fig:pssm}. A complete comparison
study of MEDUSA's PSSMs with each of the above mentioned databases
as well as Gene Ontology analysis is available on the online
supplementary website.

An added advantage of MEDUSA is that we can study the regulators
whose mRNA expression is predictive of the expression of targets.
These regulators are paired with the learned PSSMs. Of the 475
regulators (transcription factors, kinases, phosphatases and
signaling molecules) used in the study, 234 are present in the
tree. We can rank these regulators by abundance score (AS), namely
the number of times a regulator appears in the tree in
different nodes. If a regulator has a large AS, then it affects
the prediction of several target genes through several nodes.
The top 10
regulators include TPK1, USV1, AFR1, XBP1, ATG1, ETR1, SDS22,
YAP4, PDR3. TPK1 is the kinase that affects the cellular
localization of the general stress response factors MSN2/4. XBP1
is an important stress related repressor. USV1 was also identified
by Segal {\em et al.}~\cite{segal:module} to be a very important stress
response regulator. A complete analysis of the regulators as well
their association with specific motifs is available on the
supplementary website.

\section{Discussion}
We have proposed a new algorithm called MEDUSA for learning binding site
motifs together with a predictive model for gene regulation.  MEDUSA jointly
learns from promoter sequence data and multiple gene expression experiments,
together with a candidate list of putative regulators (transcription 
factors and signaling molecules), and builds motif models
whose presence in the promoter
region of a target gene, together with the
activity of regulators in an experiment,
is predictive of up/down regulation of the gene.  We can readily evaluate
the predictive accuracy of the learned motifs and regulation model on
test data, and we present results for a yeast environmental
stress response dataset that demonstrate that MEDUSA's binding site
motifs are better able to predict regulatory response on held-out experiments
than binding site sequences taken from TRANSFAC or previously published
computationally-derived PSSMs.

Popular cluster-first motif discovery strategies often require complex
or even manual preprocessing to determine suitable putative clusters of
coregulated genes.  In practice, in addition to using
gene expression profiles in the clustering algorithm, one might 
need to incorporate annotation data
or even use hand curation to properly refine the putative 
clusters \cite{hughes:alignace}.  
One must then carefully apply a standard motif discovery algorithm to
find overrepresented motifs in the promoter sequences of genes in 
each cluster, which may involve
optimizing parameters in the algorithm and 
thresholds for each of the extracted motif models.  By contrast, MEDUSA avoids
clustering and manual preprocessing altogether, and automatically 
determines PSSMs together with thresholds used for determining
PSSM hits by optimizing boosting loss.  In our experiments, MEDUSA 
learned many of the binding site
motifs associated with various environmental stress responses 
in the literature.

Recent work using the framework of probabilistic graphical models has
also presented an algorithm for learning putative binding site motifs 
in the context of building an integrated regulation model \cite{segal:learningmod}.  The graphical modeling approach is appealing due to its descriptive
nature: since the graph structure encodes how different variables are 
meant to be related, it is clear how to try to interpret the results.
The MEDUSA algorithm builds binding site motifs while producing 
a single regulation model for
all target genes without introducing conceptual subunits like ``clusters''
or ``transcriptional modules''.  
This single regulation model is arguably more biologically realistic
and can capture
combinatorial regulatory effects on overlapping sets of targets. 
The regulation model can also be interpreted as a gene regulatory network,
since the activity of regulators predicts differential expression of
targets via binding sites, although necessarily this network is
large and contains many nodes.
Nonetheless, we can use this model to
address specific biological questions,
for example 
by restricting attention to particular target genes or 
experiments \cite{kundaje:robust-geneclass}, allowing meaningful
interpretation.

One difficulty of using complex graphical models is that they require careful
training methodologies to avoid poor local optima and severe overfitting.
MEDUSA can be run ``out-of-the-box'', making it easy to reproduce
results and allowing non-specialists to apply the algorithm to new datasets.
Moreover, it is difficult to statistically validate the full structure or the
components of complex graphical
models; in the literature, most work using these models for gene regulation 
has focused on biological validation of particular features in the
graph rather than generalization measures like test loss.
MEDUSA's predictive methodology---using large-margin learning
strategies to focus on improving
generalization---produces binding site motifs that achieve good accuracy
for prediction of regulatory response on held-out experiments.  The fact
that we can easily evaluate the predictive performance of our learned
motifs and regulation model gives us a simple statistical test of confidence in
our results.
The superior
performance of MEDUSA in discovering predictive motifs is
very encouraging for applying such large-margin techniques
to analysis of expression data for as-yet unannotated genomes
and for elucidating the transcriptional regulatory mechanisms
of more complex organisms.
\appendix

\begin{small}
\section*{Acknowledgments}
AK is supported by NSF EEC-00-88001.
CW and MM are partially supported by NSF
ECS-0332479, ECS-0425850 and NIH GM36277.
CL and CW are supported
by NIH grant LM07276-02, and
CL is supported by an Award in Informatics
from the PhRMA Foundation.
\end{small}

\begin{small}
\bibliographystyle{splncs}
\bibliography{references}
\end{small}

\end{document}